\pdfoutput=1
\documentclass[]{spie}      %

\usepackage[utf8]{inputenc}
\usepackage[T1]{fontenc}
\usepackage[english]{babel}

\usepackage{amsmath,amsfonts,amssymb,mathtools}
\usepackage{bm}

\usepackage{graphicx}
\usepackage{float}
\usepackage{booktabs}
\usepackage{array}

\usepackage[colorlinks=true, allcolors=blue]{hyperref}

\newcommand{\lD}{\lambda/D}
\newcommand{\cpp}{\,\mathrm{c/p}}
\newcommand{\nm}{\,\mathrm{nm}}
\newcommand{\bG}{\mathbf{G}}
\newcommand{\bI}{\mathbf{I}}
\newcommand{\bP}{\mathbf{P}}
\newcommand{\bphi}{\boldsymbol{\phi}}
\newcommand{\GtG}{\mathbf{G}^{\!\top}\mathbf{G}}
\newcommand{\DM}{\mathrm{DM}}
\newcommand{\Var}{\mathrm{Var}}
\newcommand{\Npix}{N_{\mathrm{pix}}}

\title{Upgrade of the MITHiC testbed for phase and amplitude control in the
       focal plane}

\author{Anton Neisius\supit{a,b}, \'Elodie Choquet\supit{a},
        Vincent Chambouleyron\supit{a}, and Arthur Vigan\supit{a}
\skiplinehalf
\supit{a}Aix Marseille Univ, CNRS, CNES, LAM, Marseille, France\\
\supit{b}Institut d'Optique Graduate School, Palaiseau, France
}

\authorinfo{Further author information: (Send correspondence to A.N.)\\%
A.N.: E-mail: anton.neisius@lam.fr}

\begin{document}
\maketitle

\begin{abstract}
High-contrast imaging for the direct detection of exoplanets is limited by
stellar brightness and by wavefront errors. Coronagraphy and wavefront control
reduce these effects and enable contrast levels compatible with current
testbeds and future missions such as the Roman Space Telescope. Within this
framework, the MITHiC testbed at the Laboratoire d'Astrophysique de Marseille
is being upgraded with a new deformable mirror used to control the wavefront. It
complements the existing setup, which includes a classic Lyot coronagraph and a
spatial light modulator (SLM) enabling phase and amplitude control to optimize
the dark-hole region. We also implement the Zernike wavefront sensor (ZELDA) to
correct low-order aberrations. This contribution presents recent progress on
the MITHiC upgrade and initial results, highlighting its role as a versatile
platform for developing advanced high-contrast imaging techniques.
\end{abstract}


\section{INTRODUCTION}
\label{sec:intro}

\subsection{High-contrast imaging and wavefront control}

One of the major challenges of modern astronomy is the direct detection and
characterization of exoplanets around nearby stars. The flux ratio between a
solar-type star and a Jupiter-like planet in the visible is typically of order
$10^{9}$, and $10^{10}$ for an Earth-like planet. Reaching such contrasts is the
goal of dedicated space missions such as the Roman Space Telescope~\cite{Bailey2023, Mennesson2022, Krist2023} and the future Habitable Worlds Observatory (HWO), whose coronagraph instruments combine starlight
suppression with high-order wavefront sensing and control and focal-plane
dark-hole algorithms.

Stellar light is scattered into speckles by residual wavefront errors. In space,
free from atmospheric turbulence, the dominant limitation comes from
\emph{non-common-path aberrations} (NCPA): static or quasi-static optical
aberrations introduced after the wavefront-sensing path but before the science
focal plane. These aberrations, typically of order $10$--$50\nm$ rms, create
quasi-static speckles that evolve on timescales of minutes to hours and can mimic
planetary companions. Wavefront control is therefore split by spatial and
temporal scale: low- and mid-order modes are corrected at a low cadence (of order
a few mHz), while the high-order correction that digs the dark hole is updated
much less frequently, typically once per day.

Correcting NCPA requires a dedicated control loop operating on the residuals of
the main wavefront correction. Because these residuals are small, the associated
sensors can be highly sensitive: the unmodulated pyramid wavefront sensor and the
Zernike wavefront sensor (ZWFS) are natural
candidates~\cite{Ndiaye2013, Chambouleyron2021}. Laboratory testbeds such as
THD2~\cite{Baudoz2024} and HiCAT~\cite{Soummer2024} play a central role in
developing and validating these techniques in conditions representative of space
instruments, and MITHiC contributes to this effort.

\subsection{The MITHiC testbed}

\begin{figure}[htbp]
  \centering
  \includegraphics[width=0.9\linewidth]{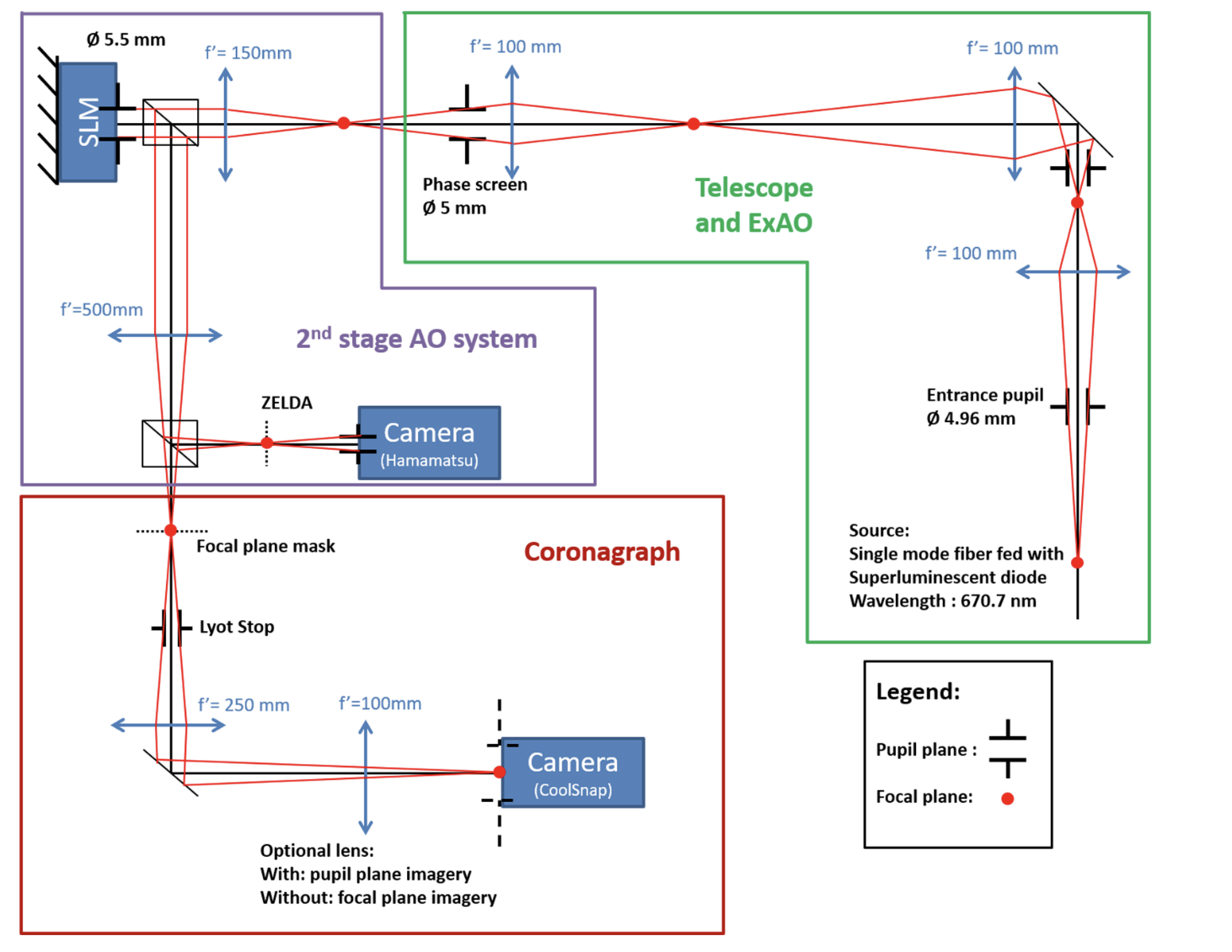}
  \caption{Optical layout of the MITHiC testbed, taken
    from~\cite{Pourcelot2022}. The beam propagates from the single-mode-fiber
    source through the SLM (pupil plane), the ZELDA wavefront sensor, the Lyot
    coronagraph, and onto the science camera.}
  \label{fig:bench}
\end{figure}

MITHiC (\emph{Marseille Imaging Testbed for High-contrast Imaging in the lab})
is a visible-light optical bench at LAM designed to test NCPA correction methods
and high-contrast coronagraphic imaging algorithms, including dark-hole
techniques such as COFFEE~\cite{Paul2013} and nonlinear dark-hole
algorithms~\cite{Leboulleux2018}. A schematic of the optical layout is shown in
Figure~\ref{fig:bench}; for a complete description of the bench, we refer the
reader to~\cite{Pourcelot2022}.

The source is a single-mode laser diode at $\lambda = 670.7\nm$ injected into a
single-mode fiber, simulating a point star. The wavefront corrector is a spatial
light modulator (SLM) providing $274\times274$ pixels across the pupil, used as
a high-resolution deformable element for NCPA correction. Wavefront sensing is
performed by a ZELDA sensor~\cite{Ndiaye2013}, consisting of a focal-plane phase
mask of $64\,\mu$m diameter combined with a pupil-plane camera, which provides a
direct pixel-by-pixel reconstruction of the optical path difference (OPD). A
science camera, operating at a gain of $g_{\mathrm{cam}} = 2.01$~e$^-$/ADU
(electrons per analog-to-digital unit), records the PSF and coronagraphic
images. Calibrated static aberrations , including segmented-mirror co-phasing
errors, are introduced by a phase screen simulating a realistic instrument.
Coronagraphic suppression is provided by a Lyot coronagraph, with a focal mask
of $2.69\,\lD$ followed by a Lyot stop. Finally, motorized stages along the $x$,
$y$, and $z$ axes allow precise positioning of the ZELDA mask and of the focal
coronagraphic mask. The main bench parameters are summarized in
Table~\ref{tab:params}.

\begin{table}[htbp]
  \centering
  \caption{Key MITHiC parameters.}
  \label{tab:params}
  \begin{tabular}{ll}
    \toprule
    Parameter & Value \\
    \midrule
    Wavelength & $\lambda = 670.7\nm$ \\
    Image size & $418 \times 418$~px \\
    Pupil mask diameter & $400$~px \\
    SLM pupil diameter & $274$~px \\
    Science camera plate scale & $12.0$~px$/(\lD)$ \\
    ZELDA exposure time & $20$~ms \\
    Science camera gain & $g_{\mathrm{cam}} = 2.01$~e$^-$/ADU \\
    Science camera frames & $N_{\mathrm{frames}} = 50$ \\
    \bottomrule
  \end{tabular}
\end{table}

\subsection{Scope of this paper}

This paper presents an upgrade of the MITHiC testbed, carried out as part of a
broader program aimed at implementing focal-plane dark-hole algorithms and,
ultimately, combining two active deformable mirrors for full amplitude-and-phase control. We report
on the characterization of the nominal bench performance and on three new
capabilities developed during this work. Section~\ref{sec:recommission}
describes the bench configuration and the characterization of its nominal
performance. Section~\ref{sec:ncpa} presents the NCPA correction loop and its
results. Section~\ref{sec:contrast} characterizes the radial contrast profile
and the photon-noise limit. Section~\ref{sec:dm} introduces the Gaussian
deformable-mirror simulation framework and the correction-floor analysis.
Section~\ref{sec:thermal} presents the thermal-stability results.
Section~\ref{sec:perspectives} outlines the roadmap toward dark-hole
implementation and future publications.

\section{BENCH CONFIGURATION AND NOMINAL PERFORMANCE}
\label{sec:recommission}

\subsection{Bench preparation}

The upgrade began with a full alignment and calibration of the bench. Each
optical element was verified and adjusted along the beam path, from the laser
source to the science camera, with particular attention paid to the pupil
conjugation between the SLM and the ZELDA camera planes. The ZELDA focal-phase
mask was then centered iteratively, using the tip/tilt and defocus response of
the sensor as a diagnostic; a lateral misalignment of $30\,\mu$m was found to
double the residual rms after NCPA correction, motivating sub-pixel positioning
accuracy. Finally, the Lyot coronagraph was aligned manually by
centering the $2.69\,\lD$ focal mask and optimizing the Lyot-stop position.

\subsection{Nominal performance}

\begin{figure}[htbp]
  \centering
  \includegraphics[width=0.35\linewidth]{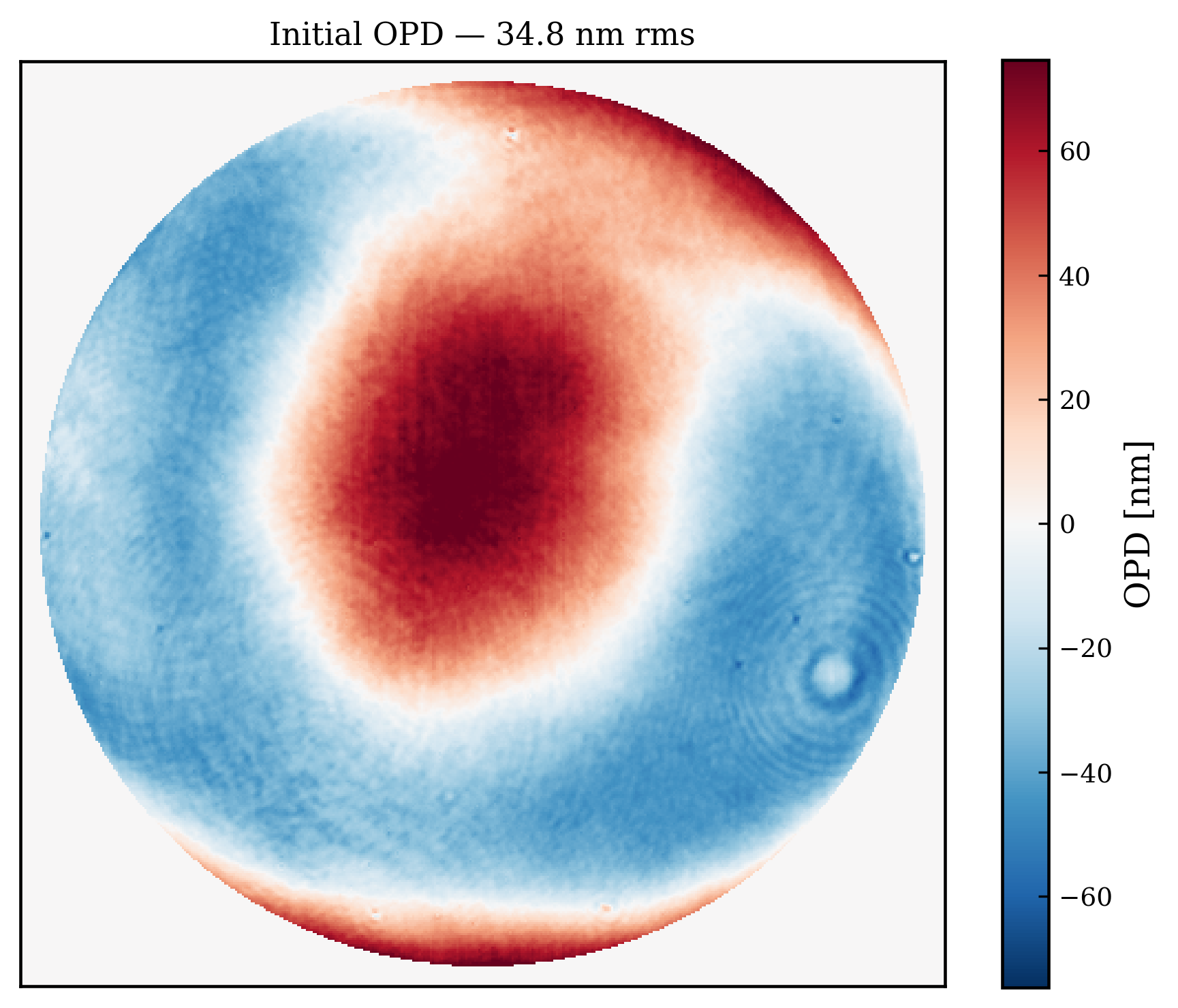}
  \caption{Initial OPD measured by ZELDA before correction ($34.8\nm$ rms).}
  \label{fig:initial_opd}
\end{figure}

With the bench aligned, the initial OPD was measured at $34.8\nm$ rms, in
agreement with the reference value reported by~\cite{Pourcelot2022}
(Figure~\ref{fig:initial_opd}). The NCPA correction pipeline
(Section~\ref{sec:ncpa}) was validated by reproducing the key figures of that
reference, confirming that the bench operates at its nominal performance level.

\section{NCPA CORRECTION WITHOUT AN INTERACTION MATRIX}
\label{sec:ncpa}

\subsection{Why no interaction matrix?}

In a classical adaptive optics system, the deformable-mirror command is computed through an \emph{interaction matrix}
$\mathbf{D} \in \mathbb{R}^{N_{\mathrm{meas}} \times N_{\mathrm{modes}}}$, which
must be calibrated by a push--pull procedure before each session. For
$N_{\mathrm{modes}}$ modes, the calibration time is
$t_{\mathrm{calib}} \approx 2\,N_{\mathrm{modes}}\,t_{\mathrm{meas}}
\approx 2\,N_{\mathrm{modes}} \times 13$~s, representing $7$~min for $20$ modes and $180$~min for $500$ modes.

On MITHiC, we deliberately operate without an interaction matrix. The
fundamental reason is that \textbf{ZELDA is a direct OPD sensor}: unlike the Shack--Hartmann sensor (which measures wavefront slopes) or the pyramid (which produces a differential signal), ZELDA provides a direct pixel-by-pixel estimate of the OPD in nanometers,
\begin{equation}
  \label{eq:zelda_opd}
  \phi(r) \approx
  \frac{I_Z(r) - I_0(r) - b^2(r)}{2\,I_0^{1/2}(r)\,b(r)},
\end{equation}
where $I_Z$ is the ZELDA image, $I_0$ the clear-pupil reference, and $b(r)$ the Fourier transform of the focal mask. The optimal correction is therefore simply $-\phi(r)$, applicable directly to the SLM without modal conversion. The SLM, with its $274\times274 \approx 75{,}000$ pixels in the pupil, provides sufficient degrees of freedom to apply this correction at full spatial resolution. The OPD reconstruction is handled by the open-source \texttt{pyZELDA} library~\cite{Vigan2019}.

This approach requires three conditions: (i) the sensor measures OPD directly in nm -- satisfied by ZELDA; (ii) the corrector can apply an arbitrary phase map at high resolution -- satisfied by the SLM; and (iii) the geometric distortion and alignment between the corrector and sensor planes -- which are both pupil planes-- are calibrated and corrected, satisfied by the affine calibration of Step~2 below.

\subsection{The integrator loop}

At each iteration $n$, the loop proceeds as follows. First, $N_{\mathrm{bg}}$
background frames, $N_{\mathrm{ref}}$ clear-pupil frames, and $N_Z$ ZELDA frames are acquired, and the OPD is reconstructed via Eq.~\eqref{eq:zelda_opd} using \texttt{pyZELDA}; piston, tip, and tilt are subtracted, as they primarily reflect Zernike mask misalignment rather than optical aberrations. The measured OPD is
then corrected for the geometric distortion between the SLM and ZELDA pupil planes, which are not perfectly conjugated: each SLM pixel is displaced by a vector $\delta\mathbf{r}$ from its nominal ZELDA position. This distortion is modeled as an affine transformation (a $2\times2$ matrix and a translation vector), calibrated once at the beginning of the session and applied at each iteration; omitting it causes the loop to diverge~\cite{Pourcelot2022}.

The distortion-corrected OPD is subsequently filtered in Fourier space,
\begin{equation}
  \phi^{(n)}_{\mathrm{filt}}(r) =
  \mathcal{F}^{-1}\!\bigl[\hat{\Pi}_{\nu_c}(\xi)\,\hat{\phi}^{(n)}(\xi)\bigr],
\end{equation}
where $\hat{\Pi}_{\nu_c}$ is a top-hat filter at cut-off frequency $\nu_c$ in cycles per pupil (c/p). This filtering serves two purposes. It prevents aliasing, the SLM pupil spans $274$ pixels, corresponding to a Nyquist frequency of $137\cpp$, and applying corrections above this frequency would cause the SLM to interpret high spatial frequencies as low-order modes and diverge , and it restricts the correction to the useful band: a sweep over $\nu_c \in \{5, 15, 30, 50, 75, 100\}\cpp$ shows that the residual rms decreases monotonically until $\nu_c = 75\cpp$ then saturates, indicating that the dominant bench aberrations lie below $75\cpp$. Finally, the SLM flat is updated as
\begin{equation}
  \hat{\phi}^{(n+1)} = \hat{\phi}^{(n)} - g\,\phi^{(n)}_{\mathrm{filt}},
  \qquad g = 1.0,
\end{equation}
with convergence guaranteed for $0 < g < 2$. In practice, ZELDA measurement noise ($\sigma_{\mathrm{ZELDA}} \approx 2\nm$ rms) and mild SLM nonlinearity beyond $|\phi| > 30\nm$ require $10$--$15$ iterations to converge.
The core pipeline comprises four sequential steps. Step~0 measures the initial OPD with the SLM set to zero. Step~1 applies a low-order pre-compensation at $5\cpp$ without distortion correction, bringing $|\phi| \lesssim 30\nm$ within the ZELDA linear range. Step~2 calibrates the geometric distortion between the SLM and ZELDA. Step~3 runs the full NCPA correction at $75\cpp$ with distortion correction, acquiring PSF images during iterations.

As a final verification of the calibration, we apply a known segmented-mirror co-phasing pattern in the optical path. This pattern is produced by the SILIOS phase screen, which emulates the segmented-pupil geometry of a realistic instrument by introducing controlled piston steps at the segment boundaries. Starting from the NCPA flat and repositioning the phase screen accordingly, we check that the loop recovers and corrects this injected pattern as expected, confirming that the distortion calibration and the correction loop behave correctly on a structured, "instrument-like" input.

\subsection{Results}

\begin{figure}[htbp]
  \centering
  \includegraphics[width=0.75\linewidth]{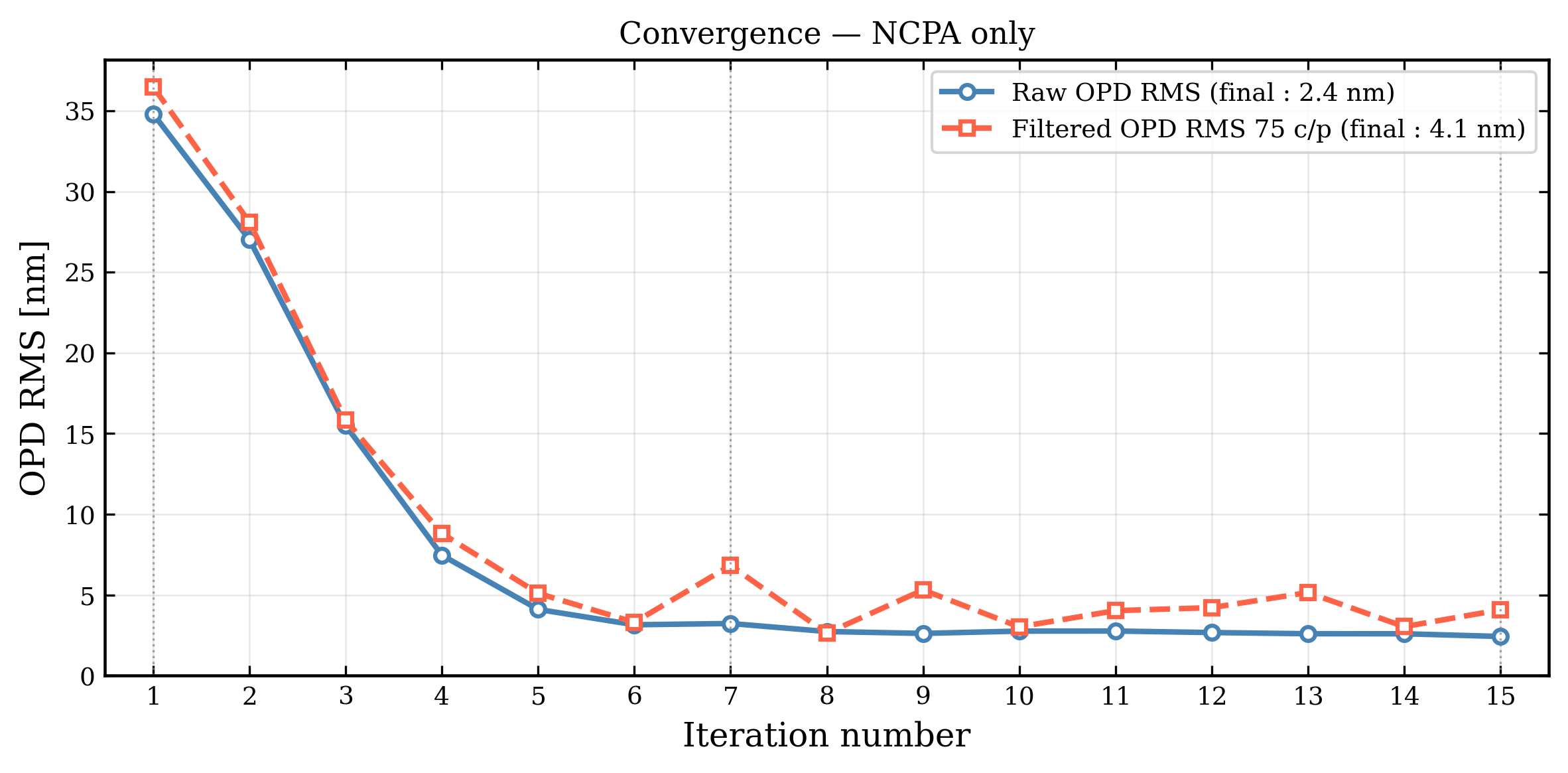}
  \caption{NCPA convergence: raw OPD rms (blue) and filtered OPD rms (red,
    $75\cpp$) versus iteration number. The final raw rms is $2.4\nm$. The
    filtered OPD rms remains slightly higher than the raw value because of
    high-order frequencies that are not corrected, attributable to a small
    residual misalignment.}
  \label{fig:ncpa_conv}
\end{figure}

The pipeline results are summarized in Table~\ref{tab:ncpa_results}, and the convergence of the NCPA loop is shown in Figure~\ref{fig:ncpa_conv}.

\begin{table}[htbp]
  \centering
  \caption{NCPA pipeline results ($15$ iterations, $g = 1.0$, $75\cpp$).}
  \label{tab:ncpa_results}
  \begin{tabular}{lcc}
    \toprule
    Step & Raw OPD rms [nm] & Filtered OPD rms [nm] \\
    \midrule
    Initial (SLM $= 0$)        & $34.8$     & $34.8$ \\
    Post pre-comp.\ $5\cpp$    & $\sim\!22$ & $\sim\!3$ \\
    Post NCPA $75\cpp$         & $2.4$      & $4.1$ \\
    Post co-phasing            & $8.6$      & --- \\
    \bottomrule
  \end{tabular}
\end{table}

The co-phasing step yields a higher residual rms than NCPA alone. This is
expected: the SILIOS phase screen introduces piston discontinuities at segment boundaries, creating high spatial frequencies at $\nu \sim D/d_{\mathrm{seg}}$ that lie above the $75\cpp$ correction bandwidth. These high-order components are not correctable by the current low-pass integrator and accumulate in the residual. A modal interaction matrix or a focal-plane electric-field conjugation algorithm would be required to address them which is precisely the motivation for the dark-hole implementation described in Section~\ref{sec:perspectives}.

\section{RADIAL CONTRAST PROFILE AND PHOTON-NOISE LIMIT}
\label{sec:contrast}

\subsection{Contrast definition}

The radial normalized intensity profile $C_\mu(r)$ is defined as the azimuthal mean of the normalized PSF intensity at angular separation $r$,
\begin{equation}
  \label{eq:contrast}
  C_\mu(r) = \frac{1}{N_{\mathrm{ring}}(r)}
  \sum_{\substack{(x,y)\,:\\ |r_{xy} - r| < dr/2}}
  \frac{I(x, y)}{I_{\mathrm{peak}}},
\end{equation}
where $r_{xy}$ is the angular distance from the PSF center in $\lD$ units and
$I_{\mathrm{peak}}$ is the off-axis reference PSF peak.

\subsection{Photon-noise normalized intensity limit}
\label{subsec:photon}

The photon-noise limit sets the fundamental physical floor on the normalized intensity measurement. We derive it here from first principles.

When the science camera measures $I_i$ ADU in pixel $i$, the corresponding number of detected photo-electrons is $N_{e,i} = g_{\mathrm{cam}}\,I_i$, where
$g_{\mathrm{cam}} = 2.01$~e$^-$/ADU is the detector gain (CoolSnap, manufacturer
specification). Photon arrival follows Poisson statistics, for which the variance
equals the mean, so the variance on the photo-electron count in pixel $i$ is
$\Var(N_{e,i}) = N_{e,i} = g_{\mathrm{cam}}\,I_i$. Expressed back in ADU, the variance per pixel is
\begin{equation}
  \label{eq:var_pixel}
  \Var(I_i) = \frac{N_{e,i}}{g_{\mathrm{cam}}^2} = \frac{I_i}{g_{\mathrm{cam}}}.
\end{equation}

Consider now the annulus of radius $r$ containing $N_{\mathrm{ring}}(r)$
statistically independent pixels, and let
$S(r) = \sum_{i \in \mathrm{ring}} I_i$ be the total signal in the ring,
proportional to the number of photons collected at that separation. Since the pixels are independent, the variances add, and
\begin{equation}
  \label{eq:var_ring}
  \Var\bigl(S(r)\bigr) = \sum_{i \in \mathrm{ring}} \Var(I_i)
  = \frac{S(r)}{g_{\mathrm{cam}}},
\end{equation}
so that the absolute photon noise on the ring sum is
$\sigma_S(r) = \sqrt{S(r)/g_{\mathrm{cam}}}$ ADU. The contrast is the ring mean normalized by the reference peak,
$NI_\mu(r) = S(r)/(N_{\mathrm{ring}}\,I_{\mathrm{peak}})$. Propagating $\sigma_S$
through this linear relation and dividing by $\sqrt{N_{\mathrm{frames}}}$ to account for the averaging of $N_{\mathrm{frames}}$ independent exposures, the
noise on the normalized intensity is
\begin{equation}
  \label{eq:sigma_contrast}
  \sigma_C(r)
  = \frac{\sigma_S(r)}{N_{\mathrm{ring}}(r)\,I_{\mathrm{peak}}\,
          \sqrt{N_{\mathrm{frames}}}}
  = \frac{1}{I_{\mathrm{peak}}}
    \sqrt{\frac{S(r)}{g_{\mathrm{cam}}\,N_{\mathrm{ring}}^2(r)\,
          N_{\mathrm{frames}}}}.
\end{equation}
Applying the $5\sigma$ detection threshold and introducing the mean ring
intensity $\bar{I}(r) = S(r)/N_{\mathrm{ring}}(r)$, the photon-noise normalized intensity
limit takes its final form,
\begin{equation}
  \label{eq:photon_noise}
  C_{\mathrm{photon}}(r) =
  \frac{5}{I_{\mathrm{peak}}}
  \sqrt{\frac{\bar{I}(r)}{g_{\mathrm{cam}}\,N_{\mathrm{ring}}(r)\,
        N_{\mathrm{frames}}}}.
\end{equation}
Because the number of pixels in a ring grows linearly with radius,
$N_{\mathrm{ring}}(r) \approx 2\pi r\,(dr)\,(\text{plate scale})^2 \propto r$, this term alone would make the limit decay as $r^{-1/2}$, appearing as a straight line of slope $-1/2$ on a log--log plot. The full radial behavior is however also set by the mean ring intensity $\bar{I}(r)$ under the square root: as the PSF
profile decreases with separation, $\bar{I}(r)$ decreases as well, so the
photon-noise limit reflects both the dilution of the noise over a growing number of pixels and the radial decay of the PSF itself.

Comparing the normalized intensity $NI(r)$ to $NI_{\mathrm{photon}}(r)$ identifies the limiting regime at each separation. Where $NI(r) \gg NI_{\mathrm{photon}}(r)$ the
image is speckle-limited and improving the NCPA correction is effective; where
$C(r) \approx C_{\mathrm{photon}}(r)$ the physical floor has been reached and
only longer integration improves the contrast, since
$C_{\mathrm{photon}} \propto 1/\sqrt{t}$. The crossover radius
$r_{\mathrm{cross}}$ separating the two regimes is a useful figure of merit for the bench.

Equation~\eqref{eq:photon_noise} assumes an unsaturated PSF peak. If
$I_{\mathrm{peak}}$ is clipped at the detector full-well capacity, both the measured normalized intensity and the photon-noise limit are underestimated; the peak ADU value must therefore be kept below
$I_{\mathrm{sat}} = 2^{N_{\mathrm{bits}}} - 1$, either by adjusting the exposure time or by inserting a neutral-density filter before acquisition.

\begin{figure}[htbp]
  \centering
  \includegraphics[width=0.85\linewidth]{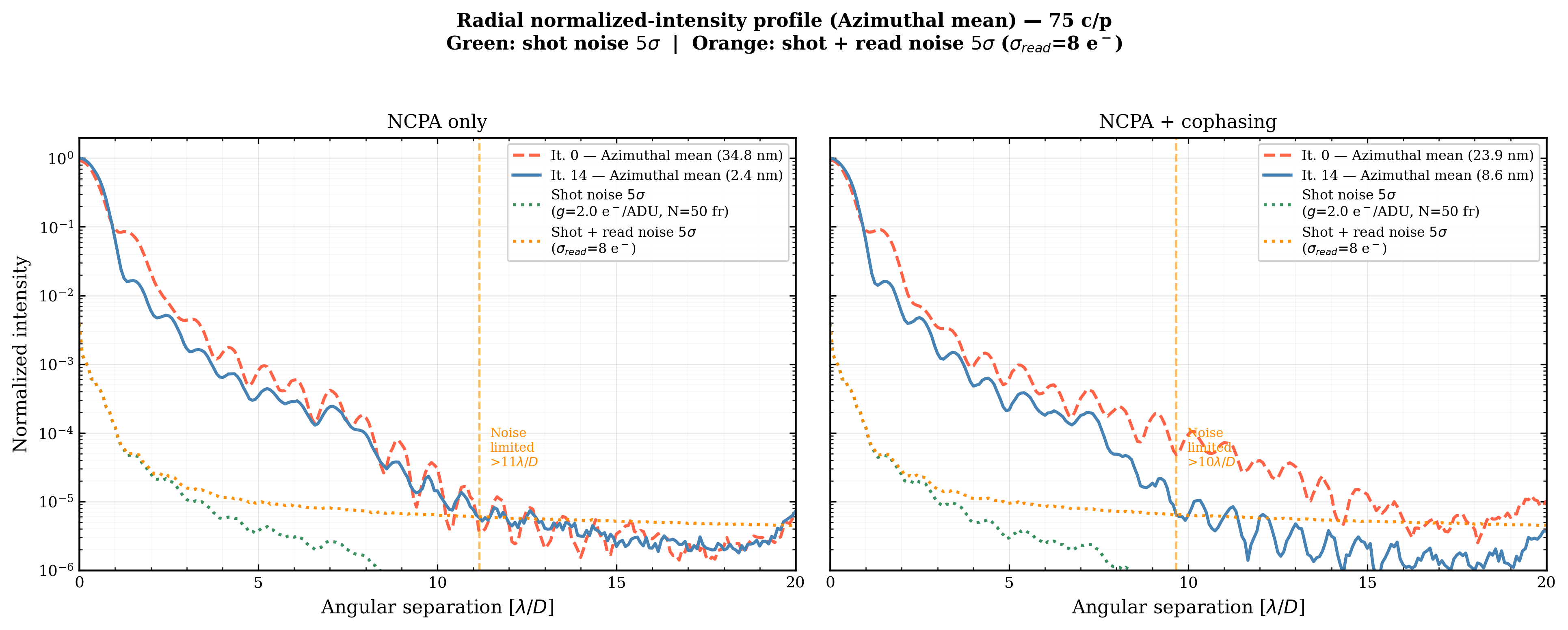}
  \caption{Radial normalized-intensity profiles before (red dashed, it.~0) and
    after (blue solid, it.~14) NCPA correction at $75\cpp$. Green dotted:
    photon-noise $5\sigma$ limit ($g_{\mathrm{cam}} = 2.01$~e$^-$/ADU,
    $N = 50$ frames). Orange dotted: photon $+$ read noise $5\sigma$ limit
    ($\sigma_{\mathrm{read}} = 8$~e$^-$). The vertical dashed line marks
    $r_{\mathrm{cross}} \approx 11\,\lD$ (left) and $10\,\lD$ (right). Left:
    NCPA only. Right: NCPA $+$ co-phasing.}
  \label{fig:contrast}
\end{figure}

After NCPA correction ($\sigma_\phi = 2.4\nm$ rms, $N_{\mathrm{frames}} = 50$,
$g_{\mathrm{cam}} = 2.01$~e$^-$/ADU), the noise-limited regime -- including the
detector read-noise contribution ($\sigma_{\mathrm{read}} = 8$~e$^-$) -- is
reached beyond $r_{\mathrm{cross}} \approx 11\,\lD$ (Figure~\ref{fig:contrast}).
Below $11\,\lD$ the contrast is dominated by residual speckles.

\section{GAUSSIAN DEFORMABLE-MIRROR SIMULATION}
\label{sec:dm}

\subsection{Motivation and relevance to dark holes}

A physical deformable mirror corrects wavefronts only within the spatial
bandwidth set by its actuator count and pitch. By restricting the SLM correction to the subspace spanned by $N^2$ Gaussian influence functions, we emulate the behavior of a real DM with $N^2$ actuators directly on the bench, without any hardware modification. This provides a controlled way to study how the achievable wavefront correction depends on the DM geometry, and it lays the groundwork for the dark-hole studies planned in future work (Section~\ref{sec:perspectives}).

\subsection{Gaussian influence-function model}

\begin{figure}[htbp]
  \centering
  \includegraphics[width=0.85\linewidth]{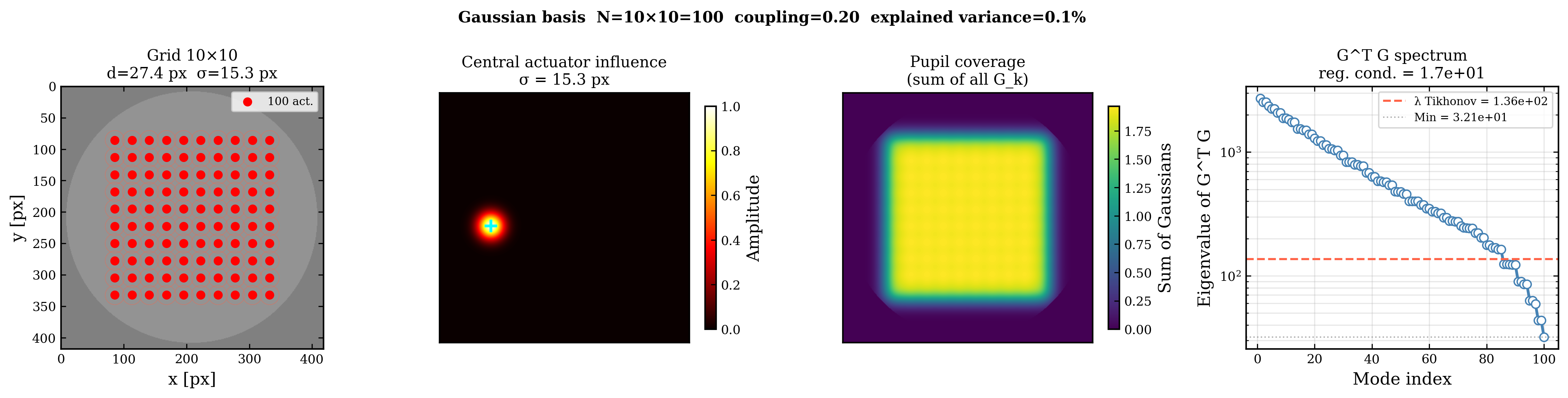}
  \caption{Simulated $10\times10$ Gaussian DM with coupling $c = 0.20$ spanning
    the full pupil: actuator grid, central influence function, total pupil
    coverage, and the eigenvalue spectrum of $\GtG$.}
  \label{fig:dm_basis}
\end{figure}

The $k$-th actuator influence function, centered at $(x_k, y_k)$, is
\begin{equation}
  G_k(x, y) = \exp\!\left(
  -\frac{(x-x_k)^2 + (y-y_k)^2}{2\sigma^2}\right),
  \qquad \sigma = \frac{d}{\sqrt{-2\ln c}},
\end{equation}
where $d$ is the actuator pitch and $c$ the inter-actuator coupling coefficient.
Stacking the $N^2$ influence functions as the columns of a matrix $\bG$, the DM
correction is the orthogonal projection of the OPD $\bphi$ onto
$\mathcal{V}_\DM = \mathrm{Im}(\bG)$,
\begin{equation}
  \bphi_\DM =
  \underbrace{\bG(\GtG + \lambda\bI)^{-1}\bG^{\!\top}}_{\bP_\lambda}\,\bphi,
\end{equation}
where $\lambda = \alpha_{\mathrm{Tikho}}\,\sigma_{\max}(\GtG)$ is the Tikhonov
regularization parameter. We compare two inversion methods, Tikhonov
regularization and truncated SVD (TSVD), the latter yielding a projector that is
idempotent to machine precision. The grid of actuators can be configured either
to span the full pupil or to cover only a fraction of it, allowing the effect of
the actuator footprint on the correction to be studied
(Figure~\ref{fig:dm_basis}).

\subsection{Correction floor}

By the Pythagorean decomposition of the orthogonal projection,
\begin{equation}
  \|\bphi\|^2 = \|\bphi_\DM\|^2 + \|(\bI - \bP)\bphi\|^2,
\end{equation}
the uncorrectable residual
$\sigma_{\mathrm{fit}} = \|(\bI-\bP)\bphi_{\mathrm{bench}}\|/\sqrt{\Npix}$ is a
fundamental floor that depends only on the bench OPD and the chosen actuator
basis -- not on the loop gain, the iteration count, or the regularization method.

\begin{figure}[htbp]
  \centering
  \includegraphics[width=0.9\linewidth]{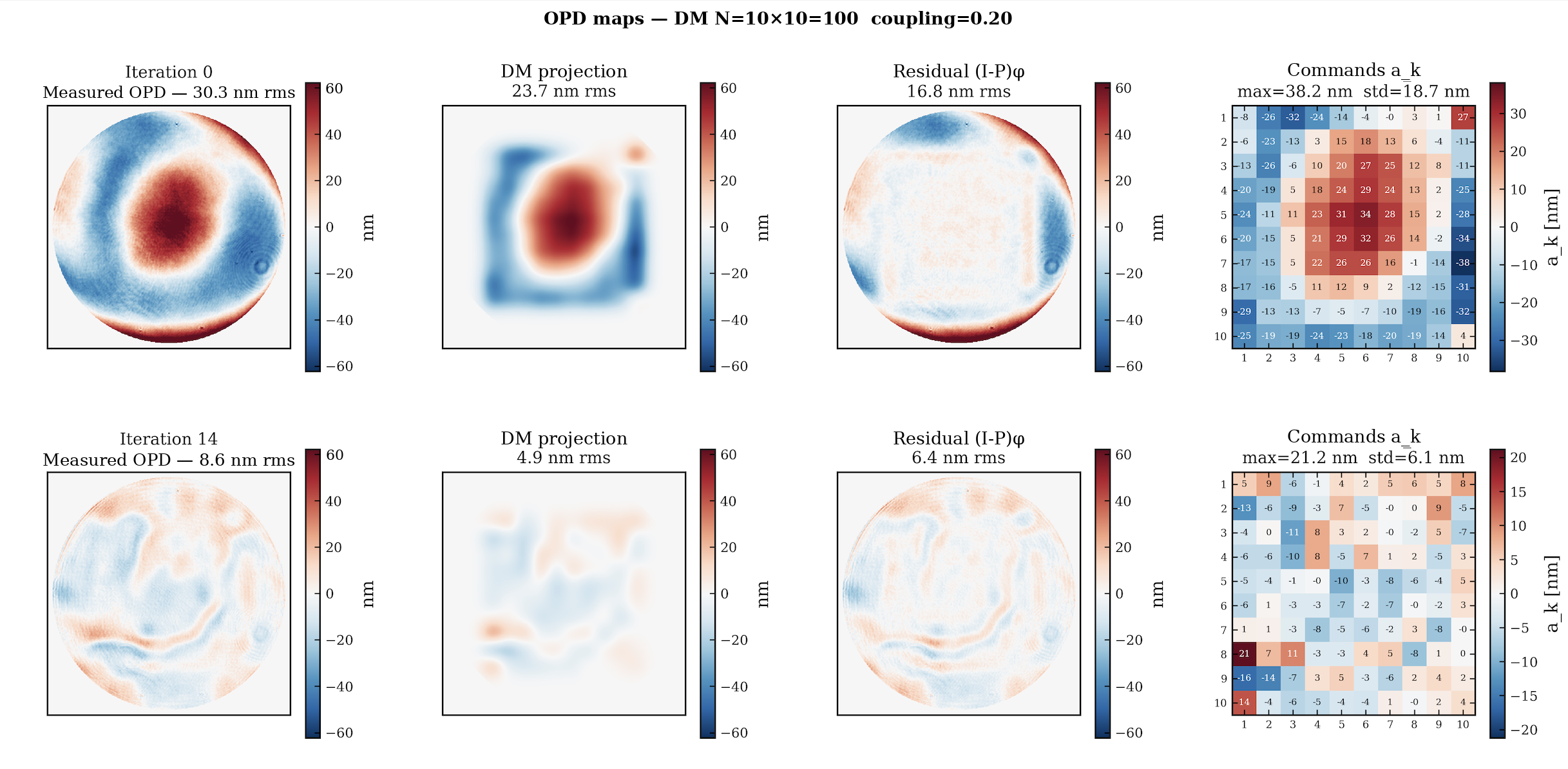}
  \caption{Closed-loop correction with the simulated Gaussian DM
    ($N = 10\times10$, $c = 0.20$): measured OPD, DM projection, residual
    $(\bI-\bP)\phi$, and reconstructed actuator commands, at the first and last
    iterations.}
  \label{fig:dm_loop}
\end{figure}

\begin{figure}[htbp]
  \centering
  \includegraphics[width=0.85\linewidth]{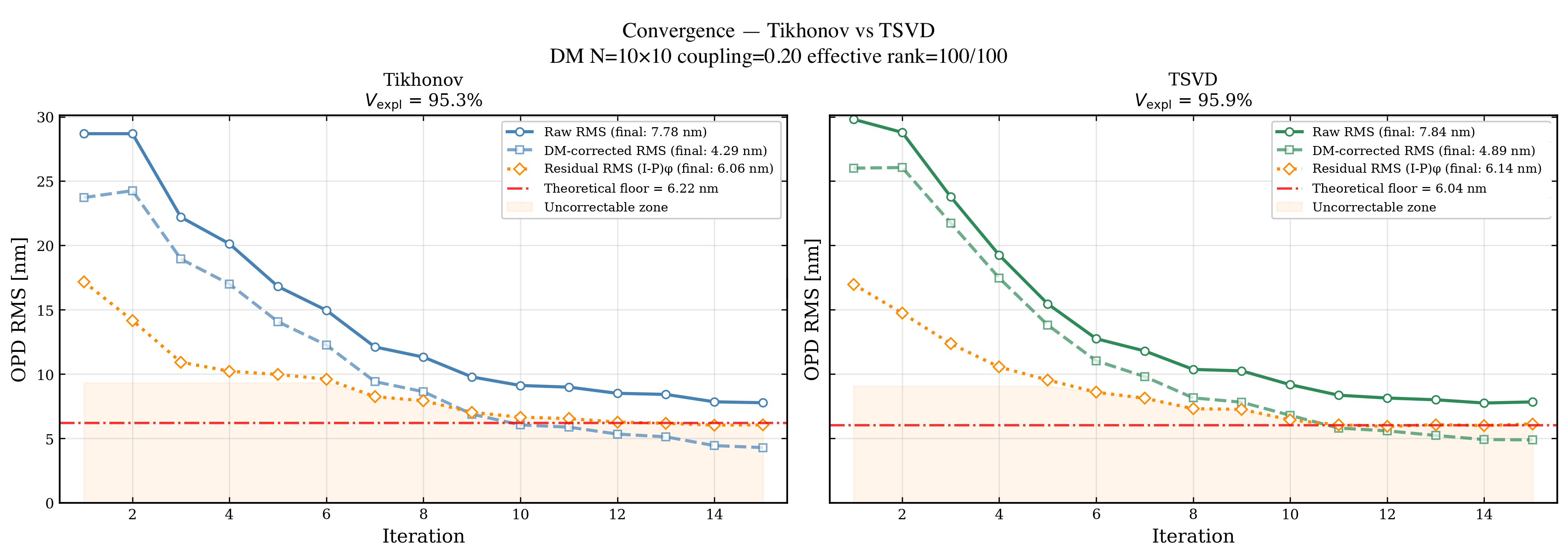}
  \caption{Closed-loop correction with a fixed Gaussian DM ($N = 10\times10$,
    $c = 0.20$). Raw OPD rms and residual $(\bI-\bP)\phi$ versus iteration
    number, for Tikhonov (blue) and TSVD (green). The initial decrease of the
    residual reflects the vanishing of the ZELDA measurement bias, not a
    reduction of the physical floor.}
  \label{fig:dm_conv}
\end{figure}

Experimentally, the residual measured by the closed loop,
$\tilde{\sigma}_{\mathrm{res}}^{(n)} = \mathrm{std}[(\bI-\bP)\bphi^{(n)}]$, is
found to decrease from $\sim\!17\nm$ to $\sim\!6\nm$ over $15$ iterations, which
at first appears to contradict the constant-floor argument above. This behavior
is a measurement bias rather than a physical effect: at iteration~$0$, the
wavefront $\|\bphi^{(0)}\| \approx 35\nm$ lies at the boundary of the ZELDA
linear range, introducing a systematic reconstruction error that inflates the
projected residual. As the loop converges and $\|\bphi^{(n)}\|$ decreases, the
bias vanishes and the measured residual converges to the true floor
(Figures~\ref{fig:dm_loop} and~\ref{fig:dm_conv}). The floor is therefore
estimated reliably as the mean over the last five iterations,
$\hat{\sigma}_{\mathrm{fit}} = \mathrm{mean}(\tilde{\sigma}^{(n)},\,n \geq N-5)$.

For a fixed DM of $N = 10\times10$ actuators with coupling $c = 0.20$ spanning
the full pupil, the closed-loop correction reduces the wavefront from $35.5\nm$
to $7.6\nm$ rms (compared with $2.4\nm$ for the unconstrained SLM), giving an
estimated floor $\hat{\sigma}_{\mathrm{fit}} \approx 6.1\nm$. The Tikhonov and
TSVD inversions give quasi-identical results at this conditioning
($\kappa = 9.2$).

\subsection{Effect of the actuator footprint}

\begin{figure}[htbp]
  \centering
  \includegraphics[width=0.7\linewidth]{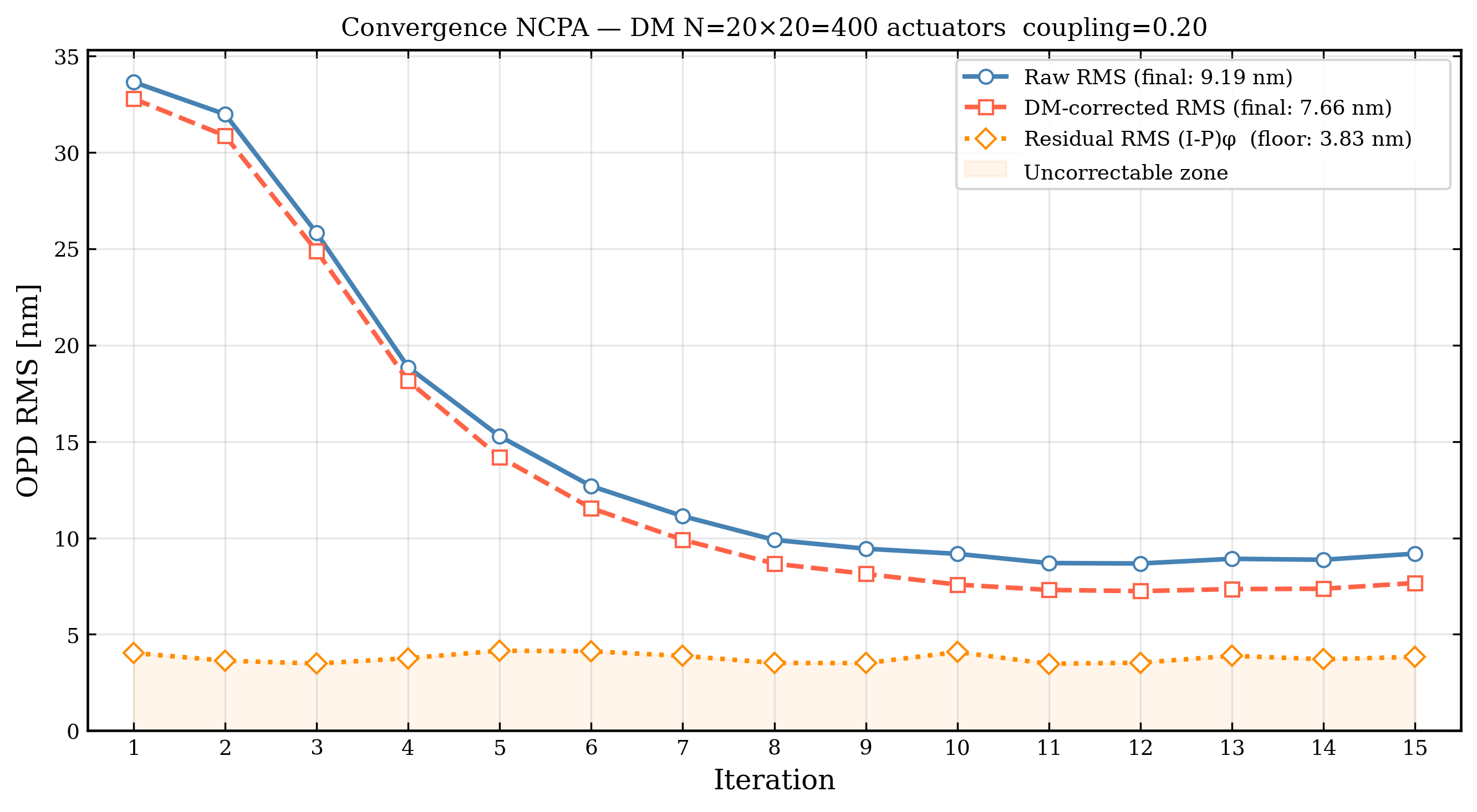}
  \caption{Closed-loop convergence for a Gaussian DM grid spanning the full
    pupil ($N = 20\times20$, $c = 0.20$): test rms, DM-corrected rms, and
    residual $(\bI-\bP)\phi$ versus iteration. The full-pupil footprint reaches
    the lowest floor.}
  \label{fig:foot_full}
\end{figure}

\begin{figure}[htbp]
  \centering
  \includegraphics[width=0.7\linewidth]{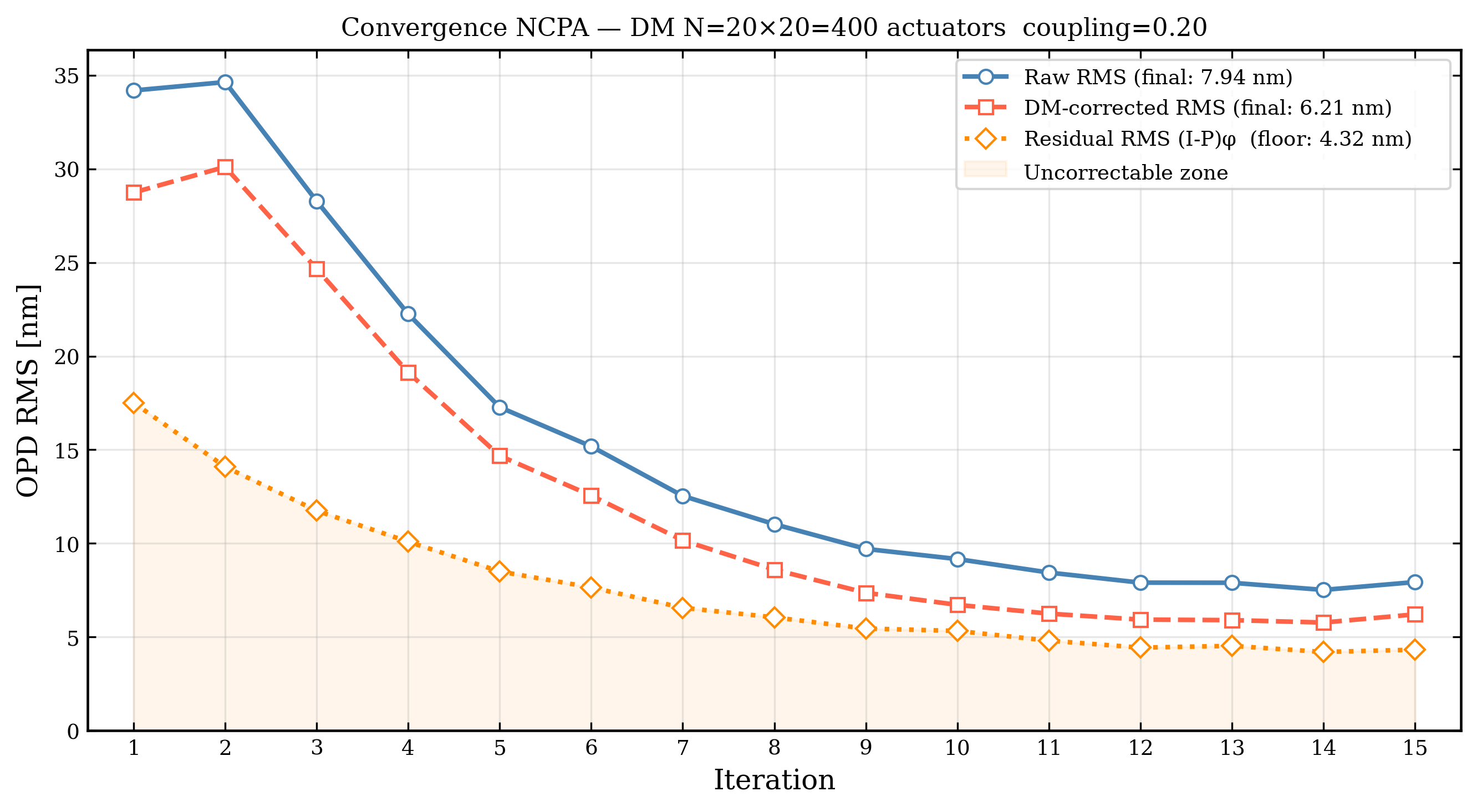}
  \caption{Same as Figure~\ref{fig:foot_full}, but for a grid covering only a
    central fraction of the pupil at fixed $(N, c)$. The uncorrected outer
    annulus raises the residual floor relative to the full-pupil case.}
  \label{fig:foot_partial}
\end{figure}

We compare two grid configurations at fixed $(N, c)$: one in which the actuator
grid spans the entire pupil, and one in which it covers only a central fraction
of it (Figures~\ref{fig:foot_full} and~\ref{fig:foot_partial}). When the grid
spans the full pupil, the influence functions reach the pupil edge and the
correction is uniform across the aperture. When the grid is smaller than the
pupil, the outer annulus is left uncorrected and contributes a structured
residual, increasing $\hat{\sigma}_{\mathrm{fit}}$. This illustrates the
trade-off between actuator pitch and pupil coverage that constrains the design of
a real DM.

\section{THERMAL STABILITY}
\label{sec:thermal}

A $12$-hour stability run records the ZELDA OPD every minute while
simultaneously monitoring bench, ZELDA-camera, and room temperatures. The
SLM-to-ZELDA distortion calibration is repeated every $30$ minutes to separate
mechanical from optical drifts.

\begin{figure}[htbp]
  \centering
  \includegraphics[width=0.95\linewidth]{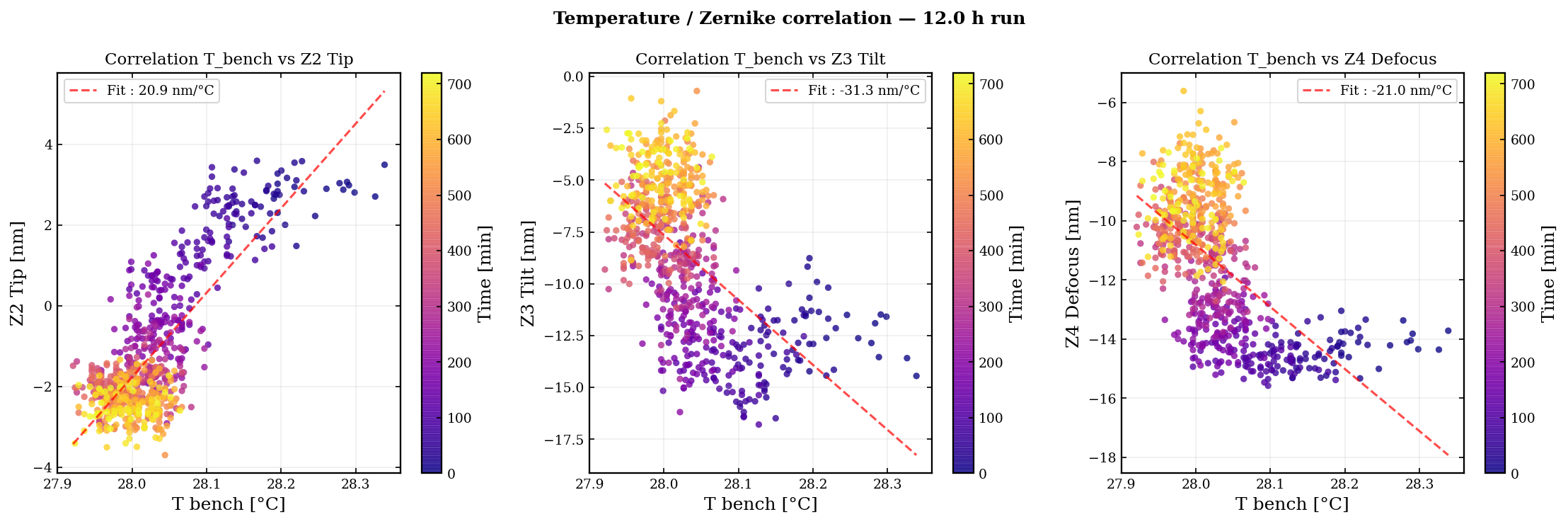}
  \caption{Zernike coefficients $Z2$--$Z4$ versus bench temperature over the
    $12$-h run. Linear fits (dashed lines) give the $\partial a_j/\partial T$
    values quoted in the text. Color encodes elapsed time.}
  \label{fig:thermal}
\end{figure}

Linear correlations between Zernike coefficients and bench temperature are
observed (Figure~\ref{fig:thermal}): $\partial a_{Z2}/\partial T =
+20.9$~nm/$^\circ$C (tip), $\partial a_{Z3}/\partial T = -31.3$~nm/$^\circ$C
(tilt), and $\partial a_{Z4}/\partial T = -21.0$~nm/$^\circ$C (defocus). For a
typical nightly temperature variation of $0.4^\circ$C, the expected tip/tilt
drift is $8$--$12\nm$ -- comparable to the NCPA correction residual. The
distortion matrix is stable to four significant figures over $12$~h, confirming
that the Zernike drifts are of thermo-mechanical origin (bench optics expansion)
rather than SLM electronic drift.

These thermal sensitivities provide the foundation for an open-loop predictive
correction: by monitoring the bench temperature, low-order aberration drifts can
be compensated without closing the ZELDA loop, which will be important for
maintaining dark-hole stability over long coronagraphic exposures.

\section{PERSPECTIVES: TOWARD DARK-HOLE IMPLEMENTATION}
\label{sec:perspectives}

\subsection{Roadmap for 2027}

The results presented here establish the quantitative reference from which
MITHiC's second development phase will proceed, along three main directions.
First, focal-plane dark-hole algorithms will be implemented, starting with
implicit electric-field conjugation (IEFC)~\cite{Haffert2023} and followed by
full electric-field conjugation (EFC). Both operate in the focal plane and
correct the complex electric field, suppressing stellar light in a defined
dark-hole region; they are the natural complement to the NCPA correction
demonstrated here, in that the integrator loop corrects the pupil-plane OPD while
EFC and IEFC optimize the focal-plane intensity directly. Second, the measured
Zernike--temperature sensitivities ($\sim\!20$--$30$~nm/$^\circ$C,
Section~\ref{sec:thermal}) will be exploited to apply open-loop predictive
corrections from temperature readings alone, maintaining dark-hole stability over
long exposures without continuous ZELDA measurements. Third, a second active component (deformable
mirror) will be added in an intermediate pupil plane to enable independent
control of amplitude and phase, allowing full electric-field control in the focal
plane.

\subsection{The correction floor as a figure of merit for dark holes}

The Gaussian DM simulation framework of Section~\ref{sec:dm} provides more than a
characterization tool: it directly addresses how dark-hole performance scales
with DM actuator count. Because the SLM allows continuous variation of $N$ and
$c$ without any hardware modification, MITHiC is uniquely positioned to map the
function $C_{\mathrm{darkhole}}(N, c)$ systematically, on the same bench, with
the same aberrations and the same algorithm, varying only the simulated DM
parameters.

This question has so far been addressed only in simulation.
\cite{Beaulieu2017} studied analytically and through Monte Carlo simulations the
dependence of dark-hole contrast on actuator count for a two-DM architecture, and
found a weak dependence at small angular separations ($\lesssim 5\,\lD$) when the
dark hole is sufficiently small. Their analysis spans several instrument design
parameters -- actuator number, DM locations, and aberration level -- but relies
entirely on numerical simulations of randomly generated optical surfaces. On the
experimental side, testbeds such as SCoOB~\cite{VanGorkom2022},
HiCAT~\cite{Soummer2024}, and THD2~\cite{Baudoz2024} have demonstrated
dark-hole contrasts of $10^{-8}$--$10^{-9}$, but always with a fixed physical DM
and without a systematic variation of the actuator count under controlled,
reproducible conditions.

To our knowledge, the relationship $C_{\mathrm{darkhole}}(N, c)$ has therefore
never been characterized experimentally on a testbed with a systematic and
continuous variation of the simulated actuator count at fixed bench conditions.
The central question is whether the dark-hole contrast tracks the theoretical
correction floor $\sigma_{\mathrm{fit}}(N, c)$ derived in Section~\ref{sec:dm}.
If it does, the floor map would provide a direct experimental design rule for
future instruments, since the optimal $(N^\star, c^\star)$ minimizing the
dark-hole contrast could be predicted from a single OPD measurement, without
running the full EFC algorithm. This is precisely where MITHiC offers a decisive
advantage: on a bench equipped with a physical DM, varying $N$ requires changing
or masking hardware and thus modifies the bench conditions, whereas emulating the
DM in software on MITHiC allows $N \in \{5, 10, 15, 20, 30\}$ actuators per side
to be explored within a single session under identical conditions. This
controlled variability is what makes a statistically significant
$C_{\mathrm{darkhole}}(N, c)$ curve measurable.

\section{CONCLUSION}
\label{sec:conclusion}

We have presented an upgrade of the MITHiC testbed together with a
characterization of its nominal performance. NCPA correction reduces the
wavefront error from $34.8\nm$ to $2.4\nm$ rms in $15$ iterations, using ZELDA
and the SLM without an interaction matrix, and reaches the noise limit
beyond $11\,\lD$.

Three new capabilities have been developed in the course of this work: a
Gaussian-influence DM simulation framework enabling a quantitative
characterization of the correction floor as a function of actuator count; a
thermal-monitoring system revealing Zernike--temperature correlations of
$20$--$30$~nm/$^\circ$C; and a rigorous photon-noise contrast-limit analysis
derived from first principles. These results are summarized in
Table~\ref{tab:summary}.

Together, they define the quantitative reference from which focal-plane dark-hole
algorithms will be implemented in 2027. The ability of MITHiC to vary the
simulated DM parameters continuously opens the way to a systematic experimental
study of dark-hole performance as a function of actuator count -- a relationship
studied analytically by~\cite{Beaulieu2017} but, to our knowledge, not yet
measured experimentally under controlled bench conditions.

\begin{table}[htbp]
  \centering
  \caption{Summary of MITHiC performance after the upgrade.}
  \label{tab:summary}
  \begin{tabular}{lcc}
    \toprule
    Experiment & Before & After \\
    \midrule
    OPD rms (bench)            & $34.8\nm$ & $2.4\nm$ (NCPA $75\cpp$) \\
    Co-phasing                 & ---       & $8.6\nm$ \\
    Photon-noise limit         & ---       & $r_{\mathrm{cross}} \approx 11\,\lD$ \\
    Simulated DM $10\times10$  & $35.5\nm$ & $7.6\nm$ (floor: $6.1\nm$) \\
    Thermal sensitivity        & ---       & $20$--$30$~nm/$^\circ$C \\
    \bottomrule
  \end{tabular}
\end{table}

\acknowledgments
A.N. thanks Rapha\"el Pourcelot for describing and explaining his previous work
on the bench, and Edvardas Vinerskas for the many scientific discussions and his
help with adaptive optics.

This project is funded by the European Union (ERC, ESCAPE, project No 101044152).
Views and opinions expressed are however those of the authors only and do not
necessarily reflect those of the European Union or the European Research Council
Executive Agency. Neither the European Union nor the granting authority can be
held responsible for them.

\bibliographystyle{spiebib}
\bibliography{main_refs}

\end{document}